# Full-waveform tomography reveals iron spin crossover in Earth's lower mantle

**Authors:** Laura Cobden[1]\*, Jingyi Zhuang[2], Wenjie Lei,[2]†,[3] Renata Wentzcovitch[2]\*, Jeannot Trampert[1], Jeroen Tromp[3]

**Affiliations:**

[1]Department of Earth Sciences, Utrecht University; 3584 CB Utrecht, The Netherlands.

[2]Department of Earth and Environmental Sciences, Columbia University; New York NY 10027, USA.

[3]Department of Geosciences, Princeton University; Princeton NJ 08544, USA

\*Corresponding authors. Email: l.j.cobden@uu.nl (general enquiries); rmw2150@columbia.edu (ab initio modelling)

†Present address

**Abstract:** Joint interpretation of bulk and shear wave speeds constrains the chemistry of the deep mantle. At all depths the diversity of wave speeds cannot be explained by an isochemical mantle. Between 1000 and 2500 km depth, hypothetical mantle models containing an electronic spin crossover in (Mg,Fe)O provide a significantly better fit to the wave-speed distributions, as well as more realistic temperatures and silica contents, than models without a spin crossover. Below 2500 km, wave speed distributions are explained by an enrichment in silica towards the core-mantle boundary. This silica enrichment may represent the fractionated remains of an ancient basal magma ocean.

**One-Sentence Summary:** Wave-speed distributions in the lower mantle require chemical heterogeneity and electronic reconfiguration in (Mg,Fe)O

Seismic tomography provides maps of the wave-speed structure inside the Earth's mantle but interpreting these maps in terms of dynamically relevant parameters such as temperature and mineralogy is a formidable task (*1, 2*). Different thermochemical parameters can have opposing effects and hence "trade off" with each other to produce a given wave-speed value (*3*). Breaking this trade-off requires a joint interpretation of multiple observables, such as compressional (*P*) and shear (*S)* wave speeds together.

While there are many different *P* and *S* wave-speed models for the lower mantle, most were obtained independently with different data sets and at different seismic frequencies, rendering a joint interpretation ineffectual. Additionally, the majority of these models were derived using classical methods which utilize only a fraction of the information available in a seismogram, and which do not capture complex wave phenomena such as diffraction. 3-D wave speeds in such tomographic models are usually expressed as linear perturbations from a reference model, rather than absolute values. In the lower mantle, with the exception of the lowermost 300-400 km (D" layer), these perturbations are small – mostly less than 1-2%. This further obfuscates a quantitative interpretation.

Tomography models derived via full-waveform inversion, that are based on fitting whole seismograms and in which the complete physics of wave propagation is accurately incorporated, provide images of the Earth's interior that are both sharper in resolution and show larger amplitude variations (*4*). Additionally, the iterative, non-linear inversion procedure directly delivers absolute wave speeds, significantly improving the constraints that can be placed on the underlying physical properties (*5*).

While the theoretical background for full-waveform inversion was developed almost 40 years ago e.g., (*6*), performing such calculations on a global scale has only just become computationally feasible (*7*).

We present a physical interpretation of a recently-published global full-waveform tomography model (*8*), GLAD-M25, between 1,000 and 2,800 km depth. GLAD-M25 constrains bulk and shear wave speeds simultaneously using the same data and over the same range of seismic frequencies and has excellent data coverage for both *P* and *S* waves traversing the lower mantle, making a joint interpretation of the two wave speeds meaningful.

The bulk wave speed, $V_\phi = \sqrt{(K/\rho)}$, is obtained through a simple combination of the compressional ($V_P = \sqrt{((K + 4G/3)/\rho)}$) and shear ($V_S = \sqrt{(G/\rho)}$) wave speeds, where $K$ is the bulk modulus (incompressibility) of the material, $G$ is the shear modulus (rigidity) and $\rho$ is the density, i.e. $V_\phi^2 = V_P^2 - 4V_S^2/3$. Interpretation of *P*-wave speed directly is complicated because it depends on both the bulk and shear moduli that are differentially influenced by mineral physics processes. Hence, the separation of the wave speeds into bulk and shear components facilitates interpretation.

Here, we study the frequency distributions of shear and bulk wave speeds as a function of depth in GLAD-M25 and infer corresponding distributions of temperature and composition which can fit both the bulk and shear wave speeds simultaneously.

We generate hundreds of thousands of models whose temperature and composition are randomly chosen from pre-defined ranges (the Prior) in a Monte-Carlo procedure. For each model, we calculate the equilibrium mineral phase assemblage via a Gibbs energy minimization and use equation-of-state modelling to compute the bulk and shear wave speeds of the assemblage. Wave speeds are adjusted for temperature-dependent anelasticity, although the effect of this correction on the wave speeds at body-wave frequencies is very small (see Supplementary Material for further details of the methods).

We consider three different priors for the lower mantle composition (Supplementary Figure S1). In the first, all models have a pyrolite composition. Pyrolite (*9*) is the hypothetical source material for mid-ocean ridge basalts (MORB), and therefore geodynamic models of mantle convection, as well as mineral physics calculations, often begin with the assumption that this is the average bulk composition of the lower mantle. The exact definition of pyrolite varies between petrological studies, so we allow minor changes in composition between pyrolite models to accommodate this uncertainty.

In our second prior, we allow extremely broad variations in chemistry; extending continuously from the ranges seen in mantle xenoliths (*10*) up to the values seen in MORBs e.g. (*11*) and chondritic Earth models (*12*). While this gives a lot of freedom in compositional possibilities for the mantle, it also includes many intermediate compositions between pyrolite and MORB that are not realistic, and it is furthermore questionable if subducted oceanic crust can be resolved at the length-scales of seismic tomography. Therefore, in our third prior we again vary the chemistry, covering the full variability seen in xenoliths and beyond, but the ranges are more restricted such that MORB-like models are excluded.

A simple, effective method to assess the relative feasibility of the three priors is to look at scatter plots of bulk versus shear wave speed at different depths. An example is shown in Figure 1 at depth intervals of 300 km. In the pyrolite models (Fig. 1a), wave-speed variations follow a narrow diagonal trend due to temperature variations, and clearly, these models cannot capture the diversity of the bulk and shear wave speeds in GLAD-M25 simultaneously at any depth (although they fit the ranges of either the bulk or the shear wave speeds in isolation).  Assuming a different fixed composition than pyrolite would shift the clouds of the models without expanding their scatter. This gives a strong indication that variations in chemistry are required to explain seismic wave speeds in the lower mantle.

With "broad" variations in chemistry it is possible to generate bulk and shear wave speeds which match the diversity of wave speeds seen in GLAD-M25 (Fig. 1b). With "restricted" variations in chemistry (Fig. 1c), this is possible at the top of the lower mantle, but with increasing depth the overlap between the synthetic models and GLAD-M25 decreases, before improving again in the D" region.

In order to improve the fit in the mid-mantle with the "restricted" prior, we require a mechanism that reduces $V_\phi$ relative to $V_S$. Both experiments and theoretical calculations have predicted that $Fe^{2+}$ in $(Mg,Fe)O$ (ferropericlase) is susceptible to a spin state change (*13, 14*). The spin state refers to the location of the 3d electrons: in the high-spin state, four electrons occupy unpaired orbitals and two are paired; in the low-spin state all six electrons are paired, thus occupying three orbitals rather than five. At low temperatures the transition takes place abruptly, but along a lower mantle geotherm, transition from the high-spin (HS) to low-spin (LS) state is expected to take place over a broad depth interval (*15*), leading to a "mixed spin" (MS) state, the spin crossover region. Owing to theoretical approximations and experimental limitations, there is still some uncertainty on the exact depth onset and thickness of the iron spin crossover (ISC) region at mantle temperatures.

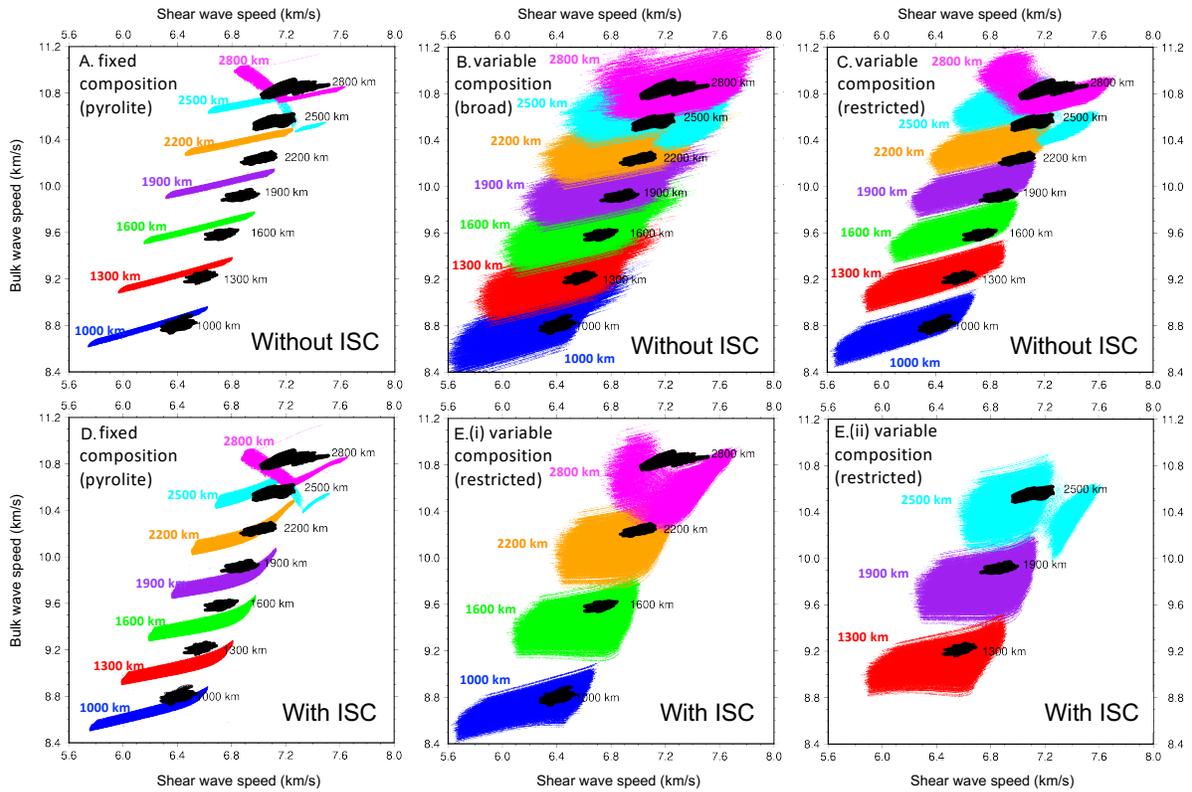

**Fig. 1. Scatter plots showing ranges of bulk and shear wave speed in seismic tomography (GLAD-M25) versus different thermochemical priors.** Black clouds are for GLAD-M25, where wave speeds have been specified at every 1 degree latitude and longitude and comprise 65,341 data points (i.e., 181 lat x 361 lon). Colored clouds are for thermochemical models: 300,000 pyrolite models and 750,000 variable composition models. Prior models are selected at random from the ranges shown in Fig S1. Top row: models without a correction for iron spin crossover (ISC) in ferropericlase. Bottom row: models with a correction for ISC in ferropericlase (**A, D**) pyrolite; (**B**) broad variable composition, (**C, E**) restricted variable composition. **E** is split into two plots (i and ii) in order to visualize overlapping ranges.

The ISC in ferropericlase is associated with a significant softening of the bulk modulus (*16, 17*), and smaller changes in the shear modulus and density (Fig. S2). This is in line with what our variable-composition models require to better fit seismic observations (Fig. 1c). However, because of the gradual and smooth nature of the ISC, and the fact that ferropericlase is expected to constitute not more than ~15-20 vol% of the bulk mineralogy, its effect on seismic properties may be subtle. The spin transition cannot be readily discerned in spherically-symmetric 1-D reference models. This is unsurprising since it would likely manifest as a change in velocity gradient with depth rather than a sharp discontinuity. The gradients in 1-D reference models are pre-determined by the parameterization choices during the inversion procedure (*18*). At the same time, averaging 3-D variations in temperature and composition into a 1-D model may mask the effect of the ISC and produce a seismic model with no physical basis (*19*). 3-D seismic tomography models are thus better suited for identifying a spin transition.

Recently, evidence for a spin transition has been suggested on the basis of differential abundances of "fast" and "slow" wave speeds between *P* and *S* wave speed tomography models (*20*). However, the tomography models used in (*20*) were derived at different seismic frequencies and resolution, with different methods and datasets and are not necessarily consistent with each other. Conclusions about the presence of the ISC were based on a "vote map" technique that extracts only the most robust and common qualitative patterns in these models.

Here, we apply a fully quantitative approach. The effect of the ISC on seismic wave speeds were recalculated using a novel non-ideal HS-LS mixing *ab initio* model for ferropericlase (see Supplementary Information for details). The non-ideal HS-LS mixing broadens considerably the ISC depth range (see Fig. S15). We investigated the effect of two approximations on the ISC pressure/temperature range: ideal vs. non-ideal HS-LS mixing and magnetic entropy. Using these new velocities we adjust the bulk modulus, shear modulus, and density of our prior models accordingly. Inclusion of the ISC effect on bulk and shear moduli expands the ranges of scatter plots (Fig. 1) significantly. For fixed composition (pyrolite) models, the scatter of bulk and shear wave speeds still does not overlap with GLAD-M25 (Fig. 1d), but for models with variable composition (Prior 3) this is now enough to fit GLAD-M25 everywhere above D" (Fig. 1e).

We can quantify the relative fit of the models with and without the ISC by applying a

Metropolis-Hastings algorithm (*21*). This procedure selects a subset of models from the prior

which can reproduce (or best fit) the frequency distributions of shear and bulk wave speeds

simultaneously, at a given depth (Fig. S3). For "restricted" chemical variations, inclusion of the

ISC improves the fit substantially between ~1,800 and 2,500 km (Fig. 2), giving us an indication

of the depth range in the mantle where the ISC is most prevalent. This improvement holds for all

four approximations made in the ferropericlase ISC modeling. In the D" region, models with an

ISC correction fit the same as, or worse than, models without this correction.

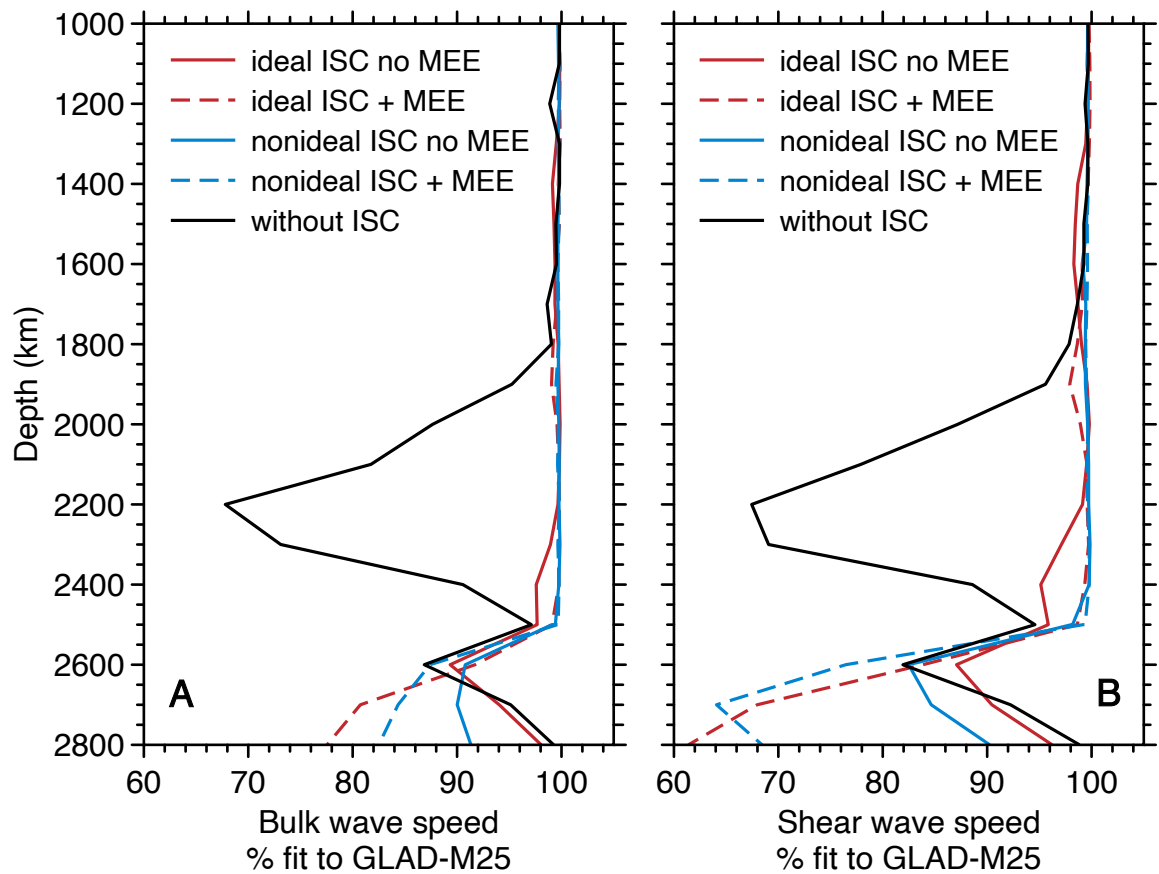

**Fig. 2. Percent fit of thermochemical models to the bulk and shear wave speed distributions in GLAD-M25 as a function of depth.** (**A**) fit to bulk wave speed distributions and (**B**) fit to shear wave speed distributions. Solid black lines are models without an ISC in ferropericlase. Colored lines show the effect of the ISC in ferropericlase using four different approximate models: ideal (red) vs. non-ideal (blue) HS-LS mixing combined with or without

magnetic entropic effects (MEE) (see details in Supplementary Materials). The results plotted here are for thermochemical models with restricted variations in chemistry (Prior 3; Fig. S1). Including the ISC significantly improves the fit between ~1,800-2,500 km depth.

By studying the physical properties of the "best fitting" models (i.e., those retained by the Metropolis-Hastings algorithm), we can further constrain the frequency distributions of temperature, bulk composition, and mineralogy which can reconstruct the wave speed distributions of GLAD-M25. The key findings are shown in Fig. 3. Although "broad" chemical variations can fit seismic observations equally well with or without an ISC (Prior 2, see Fig. S4), models without a spin correction are very cold (~800 K below a standard mantle adiabat) and they compensate for not having a spin transition with a major depletion in SiO2 in the mid-lower mantle (this manifests as a depletion in bridgmanite). The values of SiO2 (< 36 wt%) are much lower than any model proposed for the bulk mantle composition on the basis of petrological or cosmochemical arguments (*9, 12, 22-24*). Including an ISC results in mantle temperatures and silica contents which are geodynamically and geochemically more plausible, for both broad and restricted variations in chemistry.

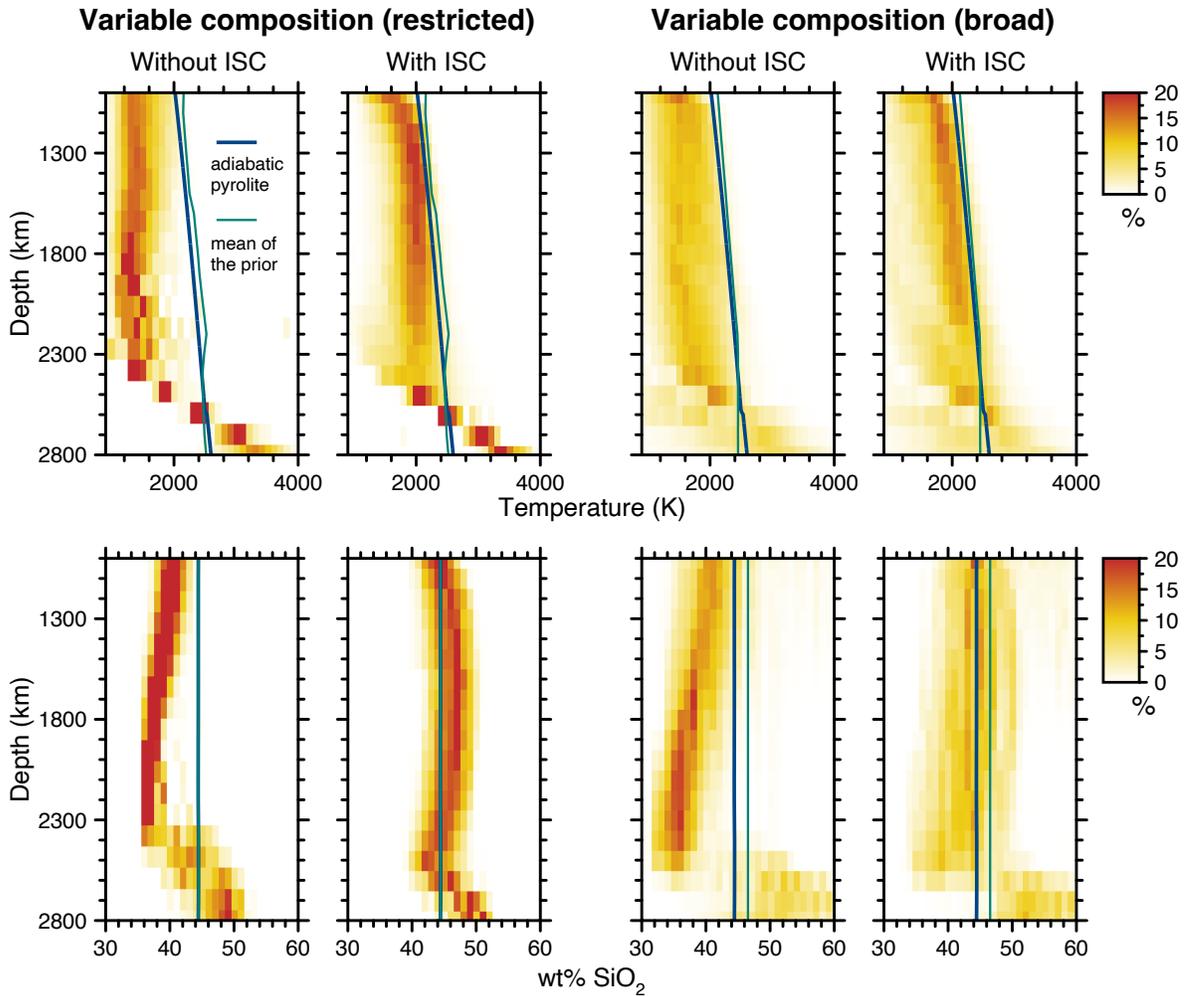

**Fig. 3. Density plots showing the distributions of temperature (top row) and silica content (bottom row) as a function of depth for the subset of thermochemical models which best fit GLAD-M25.** The darker the color, the higher the density of models (see legends). For reference, the thermal structure of an adiabat for pyrolite with a potential temperature 1,573 K is also shown (blue line) together with the mean of the prior (green line). Without a correction to the wave speeds for an ISC in ferropericlase, models are unrealistically cold and have extremely low SiO2 in the mid-mantle (especially between ~1,800-2,400 km depth). All models show and enrichment in SiO2 in the D" region, increasing towards the core-mantle boundary. Left panels are for restricted variations in chemistry (Prior 3) and right panels are for broad variations in chemistry (Prior 2). See Fig. S1 for chemical ranges of the different priors. We used results from Model-4 for the ISC (non-ideal HS-LS mixing including full magnetic entropic effects). The other three ISC models are shown in the Supplementary Materials, Fig. S5. Results for other chemical/mineralogical parameters are shown in Figs S6–S10.

In the D" region, bulk wave speed increases at a faster rate than in the overlying mantle. Our thermochemical models cater for this with an enrichment in $SiO_2$ towards the core-mantle boundary (Fig. 3), regardless of the prior ranges in chemistry or whether an ISC correction is applied. There is a greater enrichment in $SiO_2$ in the models with a broad prior, and these also fit the seismic data better (compare Fig. 2 & Fig. S4). This is because, in the absence of sufficiently $SiO_2$-rich models, the models with a restricted $SiO_2$ ranges compensate by reducing the iron content (Fig. S6) but this provides a less optimal fit to shear and bulk properties simultaneously.

Although our results are based on fitting wave speeds, density is implicitly calculated in our thermochemical models, and we can assess the plausibility of the resulting density distributions by comparing them with PREM (Fig. S11). We find that above D", all model-sets are compatible with PREM. However, in the D" region, the more Si-enriched models (from Prior 2) fit PREM better than the restricted-$SiO_2$ models (from Prior 3). The latter compensate for a lack of SiO2 – required to fit the bulk wave speed – by reducing the FeO content. This in turn reduces the density, and therefore the fit to PREM.

While the existence of a widespread ISC in Earth's lower mantle was hypothesized over 30 years ago (*13*), a seismic signal was not anticipated until recently (*17, 25*). Ascertaining the presence of the transition is important because the redistribution of electrons in $Fe^{2+}$ alters the thermal, electrical and magnetic conductivity of ferropericlase (*14, 26*), which in turn may influence the convection dynamics inside the Earth, in particular the stability of chemical piles (*27*). Previous studies e.g., (*20*) have been based on demonstrating consistency between theoretical predictions of the ISC and seismic observations. In this study, we instead quantitatively compare the fit of mantle models with and without an ISC. We demonstrate that including the elastic effects of the ISC in ferropericlase fits seismic tomography better, and that alternative explanations for the observations, namely a change in bulk chemistry by SiO2 depletion, are unfeasible. Using bulk wave speed rather than *P*-wave speed enhances the signal of the ISC (Fig. S12).

Ferric iron ($Fe^{3+}$) in bridgmanite may also experience a HS-LS crossover under lower mantle conditions and is also associated with a reduction in bulk modulus. This effect is however expected to be smaller than that in ferropericlase, and is suppressed by the presence of aluminium (*28, 29*). We can fit seismic wave speeds completely between 1,000 and 2,500 km depth by considering the ISC in ferropericlase alone, but it is possible that a similar ISC in bridgmanite may contribute to the observed signal.

Seismic tomography models depicting slab-like features traversing the whole mantle are often viewed as evidence for a chemically homogenous, well-mixed mantle. Our best-fitting mantle models require chemical heterogeneity at all depths in the lower mantle, and especially below 2,500 km. A strong enrichment in silica in the D" region may represent fractionated remnants of an ancient magma ocean (*30*) or MORB accumulation that are largely stable through geological time. These Si-rich domains could reconcile the discrepancy in Mg/Si ratio between upper mantle rocks and chondritic meteorites (*12*).

**Acknowledgments:**

**Funding:**

LC was supported by a Vidi grant from the Dutch Research Council (NWO) on grant number 016.Vidi.171.022.

RMW and JZ acknowledge funding by NSF grant EAR-2000850 and DOE grant DE-SC0019759

This research used resources of the Oak Ridge Leadership Computing Facility, which is a DOE Office of Science User Facility supported under contract DE-AC05-00OR22725 (JT).

**Author contributions:**

**JT= Jeroen Tromp; JAT= Jeannot Trampert**

    Conceptualization: LC, JT

    Methodology: LC, JZ, WL, RMW, JAT, JT

    Investigation: LC, JZ

    Formal analysis: LC

    Visualization: LC, JZ

    Resources: WL

    Funding acquisition: LC, JT, RMW

    Writing – original draft: LC, RMW

    Writing – review & editing: LC, RMW, JAT, JT

**Competing interests:** Authors declare that they have no competing interests.

**Data and materials availability:** The elastic properties of ferropericlase according to the four different theoretical models of the iron spin crossover are available at http://www.geo.uu.nl/~laura/spin2022.html . This repository also contains the data shown in the Figures.

**Supplementary Materials**

    Materials and Methods

    Fig. S1 – S16
    Table S1 – S3
    References (31 – 56)

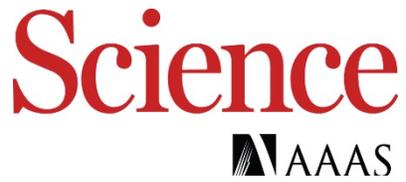

# Full-waveform tomography reveals iron spin crossover in Earth's lower mantle

Laura Cobden, Jingyi Zhuang, Wenjie Lei, Renata Wentzcovitch, Jeannot Trampert, Jeroen Tromp

Correspondence to: l.j.cobden@uu.nl (general enquiries) and rmw2150@columbia.edu (ab initio modelling)

**This PDF file includes:**

Materials and Methods
Figs. S1 to S16
Tables S1 to S3

References 31 to 56 are in the Supplementary Material only.

**Materials and Methods**

## I.  Synthetic thermochemical models

### a)  Building the prior

Bulk compositions are defined in terms of the NCFMAS system (i.e. the 6 major oxides in terrestrial rocks), and models are drawn at random from three different prior distributions (Fig. S1).

For each prior, we first prescribe a maximum and minimum value for the 6 oxides (Table S1) then select models randomly from a uniform distribution between these limits. Next, since the total weight percent of all six oxides must sum to 100%, we normalize the proportions of the six oxides to 100. For our variable composition models (Priors 2 and 3), this results in prior distributions which are non-uniform but whose peak ranges are similar to the distributions seen in xenoliths (*10*), and whose tails accommodate more extreme rock types.

For Prior 1 (pyrolite), minor variations in chemistry are included to account for variability in the formal definition of pyrolite, based on (*9, 22, 24, 31-35*). These minor variations also serve as a buffer for seismic uncertainties in GLAD-M25. We generate 1100 random pyrolite compositions. For Priors 2 and 3 (variable composition), we generate 2500 random compositions based on xenoliths (*10*), MORBs (*36*) and lower mantle models derived from chondritic and solar abundances (*22, 23*).

We analyse GLAD-M25 in depth intervals of 100 km, from 1000 km to 2800 km depth, i.e. 19 depths in total. We convert these depths into corresponding pressures using the depth-to-pressure calibration of PREM (*37*). At each of these pressures, we pick 300 temperatures at random from a uniform distribution, with a minimum temperature of 900 K and a maximum temperature at the solidus of $MgSiO_3$ (*38*). Hence, for Prior 1, this gives a total of 330,000 thermochemical models at each depth, and for Priors 2 and 3, this leads to a maximum of 750,000 models at each depth.

### b)  Calculation of seismic properties

For each of our thermochemical models, corresponding phase relations and elastic properties are calculated using *Perple_X* thermodynamic modelling software (*39*) together with the elastic parameters, equation of state, and solid solution model of (*40, 41*). Occasionally with an extreme composition or at very low temperature, models are thermodynamically unstable and are discarded from the dataset.

We first output the bulk and shear wave speed of the bulk mineral assemblage calculated directly with *Perple_X*. These wave speeds represent a thermochemical model without an iron spin crossover (ISC), and are the models plotted in Figure 1 a-c.

We then adjust the wave speeds to include the effect of an ISC in ferropericlase. We consider four different theoretical approximations for the ISC (described in Section II). We compute the change in bulk modulus, shear modulus and density as a function of temperature and iron content, relative to the high-spin (HS) state which is implicitly calculated with the database of (*40, 41*). These properties are calibrated every 100 K in temperature and 0.01 $x$Fe in ferropericlase. For each thermochemical model, we use a 2D interpolation in Python to extract the change in elastic properties in ferropericlase at the temperature and $x$Fe value for that model. With ferropericlase's properties updated for an ISC, we re-calculate the bulk and shear wave speeds (and density) of the entire mineral assemblage using a Voigt-Reuss-Hill averaging scheme.

In this procedure a small inconsistency is introduced, because the mineral phase relations are not modified for the ISC, while a redistribution of iron between ferropericlase and bridgmanite is expected (with more iron entering ferropericlase) e.g. (*14*).

While all four theoretical approximations improve the fit to GLAD-M25 in the mid-lower mantle, models which include non-ideal solid solution show more plausible thermochemical behavior after fitting to GLAD-M25 (see Figures S5-S9) and are therefore our preferred choice.

### c) Correction to wavespeeds for intrinsic anelasticity

As a last step, we apply a simple correction to adjust shear wave-speeds for the effect of temperature-dependent intrinsic anelasticity. We follow the procedure in (*42*) with the parameters given in Table S2. The effect on $S$-wave speeds is very small (< ~ 0.3%).

## II. Iron spin-crossover modeling

### a) Thermodynamic modeling

Here we model more realistically the acoustic velocities of ferropericlase (*fp*). Previously such velocities were obtained using an ideal solid-solution mixing model (*43*) and more approximate vibrational properties (*44*). There are two levels of modeling in the iron spin crossover (ISC) solid-solution: a) the MgO-FeO solid solution modeling is treated as a quasi-ideal solid-solution (*45, 46*) with end-members MgO and Mg$_{(1-x)}$Fe$_x$O ($x_{Fe}$ = 0.1875) in the high-spin (HS) of low-spin (LS) state. This level of modeling is equivalent to treating the solid solution as a Henryan solution, with an activity coefficient different from 1 but constant for small $x$; b) the Mg$_{(1-x)}$Fe$^{HS}$O and Mg$_{(1-x)}$Fe$_x^{LS}$O solution modeling with fixed $x$ and LS fraction $n = \frac{n_{LS}}{n_{HS}+n_{LS}}$ varying in the full range $0 < n < 1$. The non-ideal free energy expression is general, and its contributions were described in a recent paper (*47*):

$$G_{non-ideal}(P,T,n) = G_{ideal}(P,T,n) + G_{ex}(P,T,n). \tag{1}$$

$G_{ex}(P, T, n)$ is the excess Gibbs free energy describing the non-ideal mixing and $G_{ideal}(P, T, n)$ is

$$G_{ideal}(P, T, n) = (1 - n)G_{HS}(P, T) + nG_{LS}(P, T) + G_{mix}(P, T, n), \qquad (2)$$

where $G_{HS/LS}(P, T)$ is the Gibbs free energy of 100% HS or LS ferropericlase, i.e.,

$$G_{HS/LS}(P, T) = F_{HS/LS}^{st+vib}(V(P), T, n) + PV_{HS/LS} + G_{mag}(P, T, n). \qquad (3)$$

$P_{HS/LS} = -\frac{\partial F_{HS/LS}^{st+vib}}{\partial V}$ and $F_{HS/LS}^{st+vib}$ is described within the quasiharmonic approximation (QHA)

$$F^{st+vib}(V, T, n) = E^{st}(V, n) + F^{vib}(V, T, n). \qquad (4)$$

$G_{mix}(P, T, n)$ is the ideal free energy of mixing:

$$G_{mix}(P, T, n) = -TS_{conf}(T, n) = -k_B x_{Fe}[n \ln n + (1 - n)\ln(1 - n)], \qquad (5)$$

And

$$G_{mag}(P, T, n) = -TS_{mag}(T, n) = -k_B T x_{Fe}(1 - n)\ln[m(2S + 1)]. \qquad (6)$$

$G_{mag}(P, T, n)$ is non-zero for the HS state only. Eq. (6) assumes no exchange interaction between iron ions (no spin-spin correlations) and corresponds to the atomic limit, where $m = 3$ is the minority electron orbital degeneracy in the HS state and S = 2 is the total spin of iron in the HS state. Eq. (6) gives the maximum magnetic entropy allowed, which is a good approximation in the limit of small $x$, where Fe-Fe distances are large. For large $x$, Fe-Fe distances are small, and exchange interaction may induce magnetic ordering, decreasing $S_{mag}$. $fp$ with $x_{Fe} \leq 0.2$ may still be treated well in the atomic limit (as paramagnetic impurities), but here we inspect the effect of two limits of $S_{mag}$ on the spin-crossover: the maximum value given by Eq. (6) and the minimum value, i.e., $S_{mag} = 0$, as in a diamagnetic insulating state.

Putting all these ingredients together, we minimize $G_{non-ideal}(P, T, n)$ w.r.t $n$ to obtain the equilibrium $n(P, T)$, i.e., the solution of

$$\Delta G_{LS-HS}(P, T) + \frac{\partial G_{ex}(P,T,n)}{\partial n} + k_B T x_{Fe} \ln\left[\frac{n}{1-n}\big(m(2S + 1)\big)\right] = 0. \qquad (8)$$

In the absence of $G_{ex}$, the solution is

$$n = \frac{1}{1+m(2S+1)exp\left[\frac{\Delta G_{LS-HS}(P,T)}{k_B Tx}\right]}, \tag{9}$$

where $\Delta G_{LS-HS}(P,T) = G_{LS}(P,T) - G_{HS}(P,T)$. For non-vanishing $G_{ex}$, Eq. (6) needs to be solved numerically. Here we include only the static part of $G_{ex}(P,T,n)$, i.e., $H_{ex}(P,n) = E_{ex}^{st}(V,n) + P_{ex}(V,n)V$ and assume $G_{ex}^{mag} = F_{ex}^{vib} = 0$. This is an excellent approximation. We use a 3$^{rd}$ order polynomial to fit $H_{ex}(P,n)$ with the boundary conditions $H_{ex}(n=0) = 0$ and $H_{ex}(n=1) = 0$ at each pressure:

$$H_{ex}(V,n) = an^3 + bn^2 - (a+b)n \tag{10}$$

which produces

$$\frac{\partial H_{ex}(n)}{\partial n} = 3an^2 + 2bn - (a+b). \tag{11}$$

After obtaining $H_{ex}(V,n)$ (see below), we fit Eq. (10) at each volume and obtain $a(V)$ and $b(V)$. They are as used as in Eq. (11) and replace in Eq. (8), resulting in:

$$\Delta G_{LS-HS} + 3an^2 + 2bn - (a+b) + k_B Tx_{Fe} \ln\left[\frac{n}{1-n}\left(m(2S+1)\right)\right] = 0 \tag{12}$$

Eq. (12) is then solved numerically for $n$ at each $P,T$. This procedure was followed for $x_{Fe}$=0.1875.

Next, we obtain $H_{ex}(P,T,n)$:

$$H_{ex}(P,n) = E_{ex}^{st}(V,n) + P_{ex}^{st}(V,n)V \tag{13}$$

where $P_{ex}^{st}(V,n) = -\frac{\partial E_{ex}^{st}}{\partial V}$. The first step in this procedure consisted in obtaining $E_{ex}^{st}(V,n)$.

For 8 different volumes, $E_{ex}^{st}$ was obtained by carrying out *ab initio* calculations on a 64-atoms supercell with 6 Fe, 26 Mg, and 32 O ions. $n$ varied from 0 to 1, in steps of $\frac{1}{6}$. The supercell Mg/Fe configuration maximized iron-iron distances. The possible HS-LS iron configurations are listed in Table S3. A total of 51 HS-LS configurations are involved but only 10 with different multiplicities are inequivalent.

A typical example of the type of results we produce is seen in Fig. S13.

Fig. S13 shows $E - E_0$ where

$$E = E^{st}(V,T,n) = \frac{1}{N_n} \sum_{i=1}^{N_n} m_i^n E_i(n)\, e^{-\frac{E_i(n)}{k_B T}}, \tag{14}$$

with $m_i^n$ is the multiplicity of the $i^{th}$ inequivalent configuration with LS-fraction $n$, $N_n = \sum_{i=1}^{N_n} m_i^n$ is the total number of HS-LS configurations for $n$, $E_0 = E_{HS}(V)$ and $E_{ideal}(V,n) = (1-n)E_{HS}(V) + nE_{LS}(V)$ (blue symbols in Fig. S13). As seen, there is an insignificant temperature dependence in $E^{st}(V,T,n)$ which is rightly disregarded.

Fig. S14 shows $H_{ex}(P,n)$ at different pressures fit to a $3^{rd}$ order polynomial in $n$ as indicated in Eq. (10).

b) Ab initio calculations

Self-consistent LDA+U$_{sc}$ calculations were performed using the Quantum ESPRESSO code. The projector-augmented wave (PAW) data sets from the PSlibrary (*48*). A kinetic-energy cutoff of 100 Ry for wave functions and 600 Ry for spin-charge density and potentials were used. In all cases, atomic orbitals were used to construct occupation matrices and projectors in the LDA+ U$_{sc}$ scheme. The Hubbard parameter $U$ on Fe-3d states was computed using density-functional perturbation theory (*49*). A cubic supercell with 64 atoms was constructed, i.e., (Fe$_x$Mg$_{1-x}$)$_{32}$O$_{32}$, with x = 0.1875. The $2 \times 2 \times 2$ **k**-point mesh was used for Brillouin zone integration. Structure optimization was performed by relaxing atomic positions with a force convergence threshold of 0.01 eV/ Å. The convergence threshold of all self-consistent field (SCF) calculations was $1 \times 10^{-9}$ Ry and for DFPT calculations of U$_{sc}$ was $1 \times 10^{-6}$ Ry. Phonon calculations were performed using the finite-displacement method and the Phonopy code (*50*) with force constants computed with Quantum ESPRESSO. Vibrational density of states (VDOSs) were obtained using a q-point $20 \times 20 \times 20$ mesh. The vibrational contribution to the free energy was calculated using the quasiharmonic approximation with the qha code (*51*). More details on these *ab initio* calculations can be found in (*47*).

Here we inspect results from four thermodynamic models: a) ideal HS/LS mixing with magnetic entropy effect (MEE) ($G_{mag} = 0$) , b) non-ideal mixing with MEE ($G_{mag}$ given by Eq. (6)), c) ideal HS/LS mixing without MEE, b) non-ideal mixing without MEE. The four diagrams for $n(P,T)$ for x = 0.1875 are shown in Fig. S15.

For ideal or non-ideal ISC modeling, the inclusion of MEE decreases the slope of the ISC. With MEE, the crossover pressure range agrees better with data from Komabayashi et al. (*52*) on a sample with x = 0.19. Without MEE, the slope of the ISC agrees better with Lin et al. (*53*) data on a sample with $x$ = 0.25. This sample showed antiferromagnetic correlations at low temperatures, consistent with Fe-Fe exchange interaction with larger $x$, and lower $S_{mag}$.

The 300 K compression curves for these four models are shown in Fig. S16 below. The inclusion or exclusion of $G_{mag}$ in the calculation is not visible at 300 K for the non-ideal mixing model. MEE is distinguishable in the slope of the ISC only, for ideal or non-dial solution modeling.

*c)* <u>Thermoelasticity</u> <u>calculations</u>

The formalism for thermoelasticity with a spin crossover is described in (*43*). The components of the compliance tensor in the mixed spin (MS) state are written as:

$$S_{ij}(n)V(n) = nS_{ij}^{LS}V^{LS} + (1-n)S_{ij}^{HS}V^{HS} - \frac{1}{9}\alpha_{ij}(V^{LS} - V^{HS})\frac{\partial n}{\partial p}\Big|_{P,T}. \qquad (15)$$

All quantities in Eq. (15) are functions of pressure and temperature, e.g., $V(n) = V(P,T,n)$ or $S_{ij}^{LS} = S_{ij}^{LS}(P,T)$. For this cubic system, $\alpha_{11} = \alpha_{12} = 1$, $\alpha_{44} = 0$. The $S_{ij}^{HS/LS}(P,T)$ are obtained using by inverting the elastic tensor, $C_{ij}^{HS/LS}(P,T)$ calculated with the cij code (*54*). The compliance tensor, $S_{ij}(P,T,n)$, is then inverted to give $C_{ij}(P,T,n)$.

Bulk, $K(P,T,n)$, and shear, $G(P,T,n)$, elastic moduli can be determined from the elastic constant $C_{ij}(P,T,n)$ using the Voigt-Reuss-Hill (VRH) averaging scheme. The Voigt average assumes that strain is uniform throughout the system (upper bound). For a polycrystalline system, they are:

$$K_V = \frac{1}{9}[(C_{11} + C_{22} + C_{33}) + 2(C_{12} + C_{23} + C_{13})] \qquad (16a)$$

$$G_V = \frac{1}{15}[(C_{11} + C_{22} + C_{33}) - (C_{12} + C_{23} + C_{13}) + 3(C_{44} + C_{55} + C_{66})] \qquad (16b)$$

The Reuss bound assumes uniform stress and can be computed as

$$K_R = \frac{1}{[(S_{11} + S_{22} + S_{33}) + 2(S_{12} + S_{23} + S_{13})]} \qquad (17a)$$

$$G_R = \frac{15}{[4(S_{11} + S_{22} + S_{33})) - 4(S_{12} + S_{23} + S_{13}) + 3(S_{44} + S_{55} + S_{66})]} \qquad (17b)$$

The arithmetic average of the Voigt and Reuss bounds is the Hill average. Thus, the VRH average of the elastic moduli are

$$K_{VRH} = \frac{K_V + K_R}{2} \qquad (18a)$$

$$G_{VRH} = \frac{G_V + G_R}{2} \qquad (18b)$$

It is implicit that all quantities ($M$) above are functions of pressure, temperature, and $n$, i.e., $M(P,T,n)$, for a particular $x$. Such elastic properties ($M(P,T,x,n)$) were calculated for $x = 0$ and $x = 0.1875$ and then linearly interpolate/extrapolated for $0 \leq x \leq 0.25$.

$K_{VRH}(P,T,x,n)$, $G_{VRH}(P,T,x,n)$, and the density $\rho(P,T,x,n)$ for these four models are offered as downloadable files (*55*).

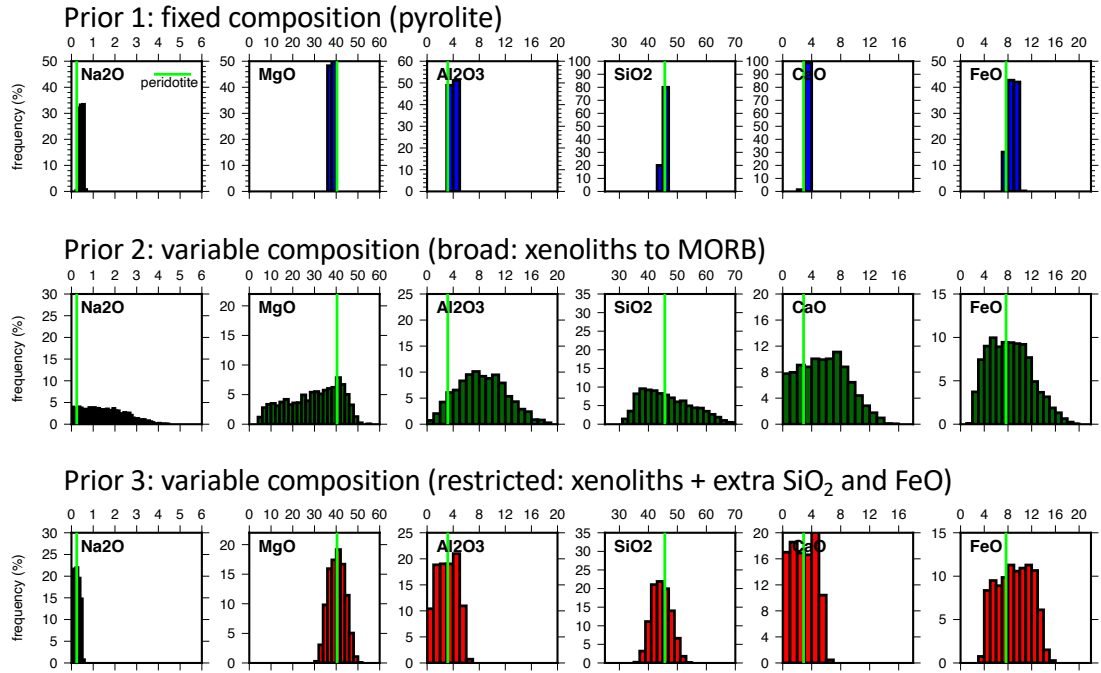

**Fig. S1.**

Frequency distributions showing the ranges of bulk composition (in wt %) for three different priors. The pale green vertical line shows the values for peridotite (*56*) for reference.

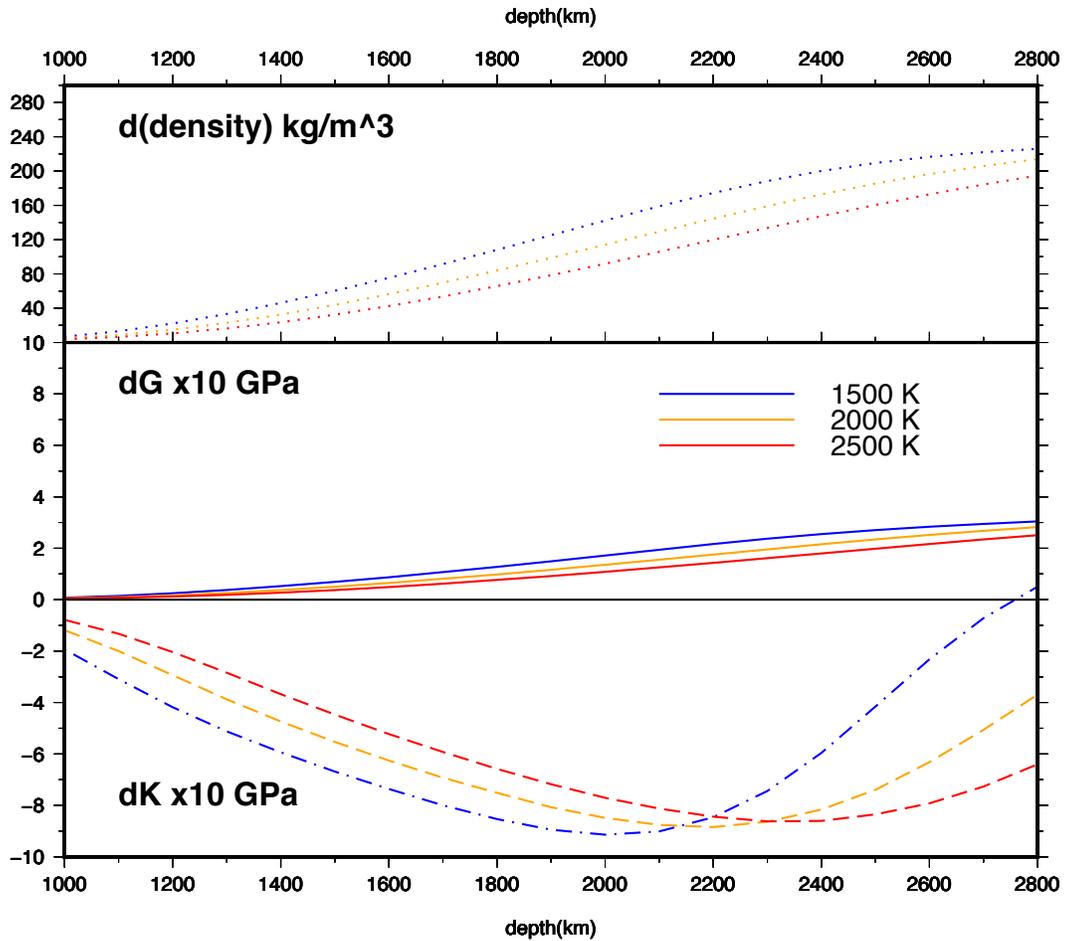

**Fig. S2.**
Change in density, shear modulus and bulk modulus as a function of depth along 3 different isotherms due to high-to-low spin transition in ferropericlase, including the effects of non-ideal solid solution and magnetic entropy.

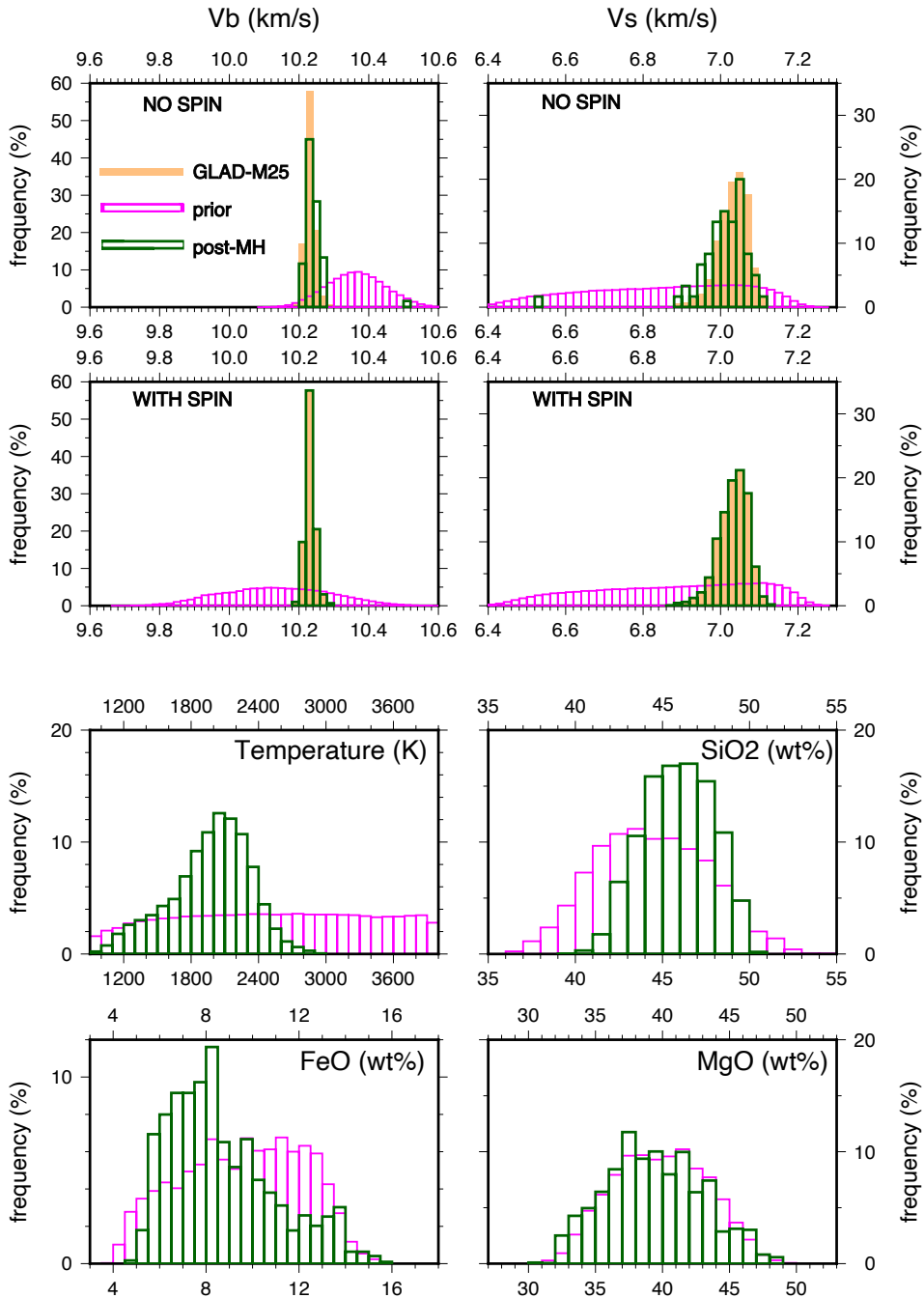

**Fig. S3.**

Illustration of application of Metropolis-Hastings (MH) algorithm at 2200 km depth. Top two rows: Yellow histograms show the frequency distributions of bulk and shear wave speed in GLAD-M25. Pink histograms show the wave-speed distributions of the prior (Prior 3, Fig. S1), with and without inclusion of effect of a spin transition in ferropericlase. Green histograms show the best fit to GLAD-M25 after applying MH. The degree of overlap between the yellow and green histograms is used to quantify the fit, as plotted in Figure 2. Clearly the fit is better with a spin transition that without. Bottom two rows: distributions of temperature, FeO, SiO2 and MgO in Prior 3 (pink histograms) versus the remaining subset of models after applying MH (green histograms).

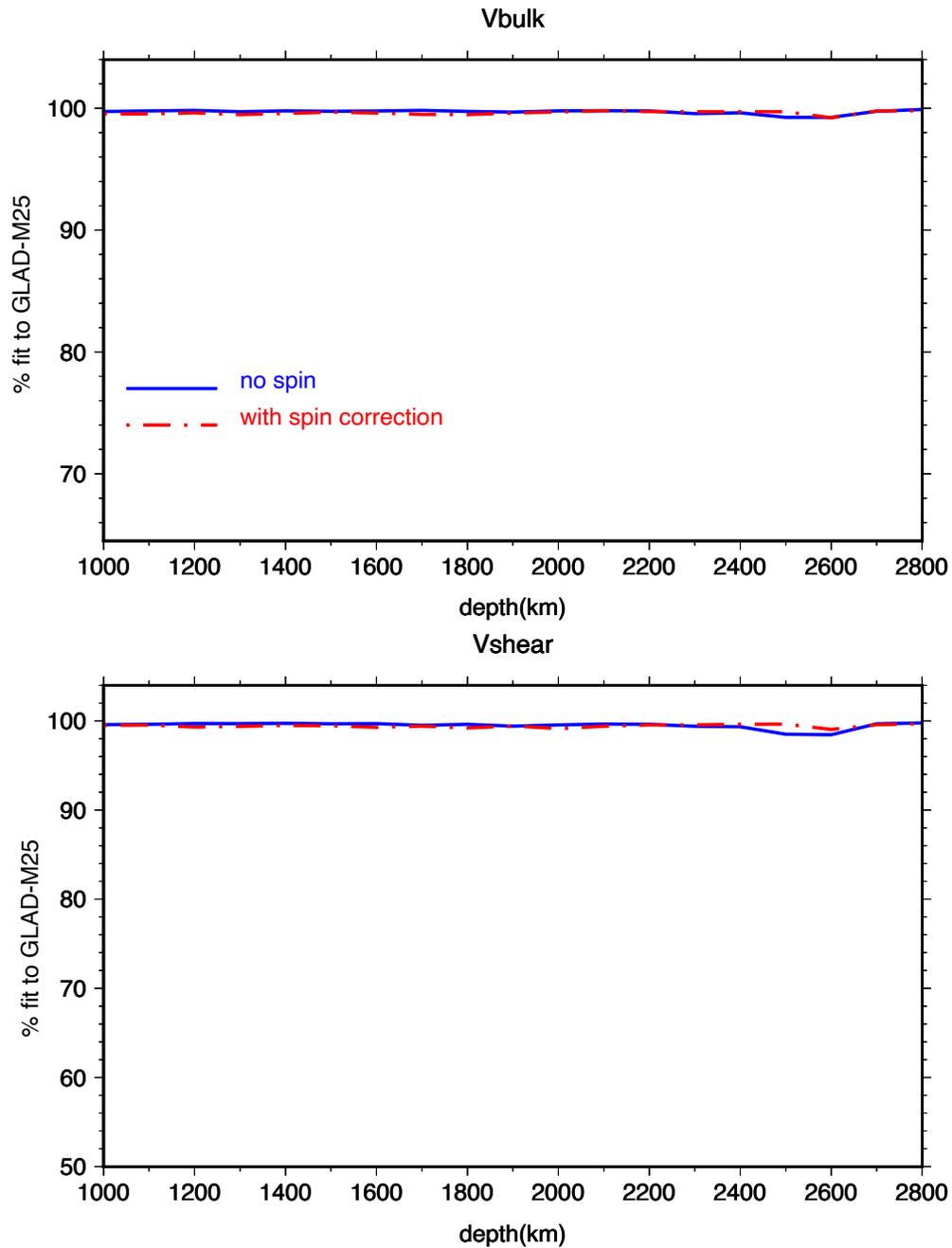

**Fig. S4.**
Comparison of fits to GLAD-M25 for Prior 2, with (red dashed line) and without (blue line) a spin transition. For clarity, results are shown for just one spin model: non-ideal solid solution with magnetic entropy. This spin model provides the most plausible temperature and composition gradients (see Figures S5-S9).

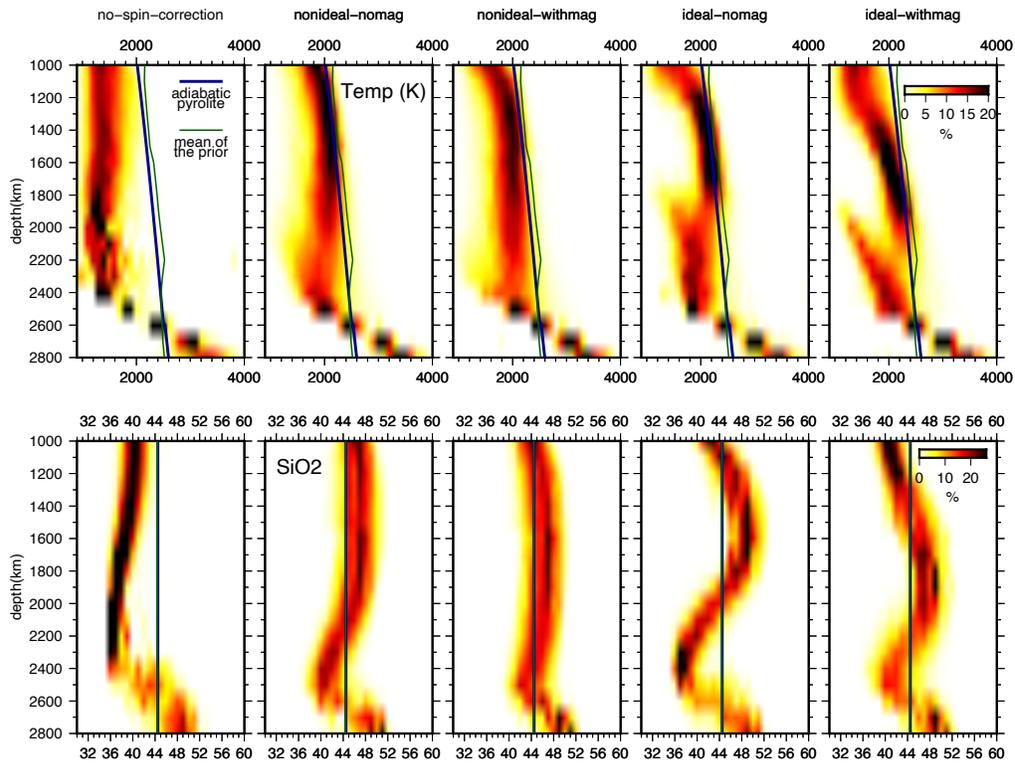

**Fig. S5.**

Density plots showing distributions of temperature (top row) and wt% SiO₂ (bottom row) as a function of depth, for models with variable, but restricted, chemical composition (Prior 3, Fig. S1). Adiabatic pyrolite with potential temperature 1573 K (blue line) and mean of the prior (green line) are shown for comparison. On the left, models without a correction to the wavespeeds for spin transition. These models are both cold and very Si-poor in the mid-mantle. Other four columns show the results for 4 different spin corrections. Models which include non-ideal solid solution give more reasonable temperature and compositional gradients, in particular the model with both non-ideal solid solution and a correction for magnetic entropy.

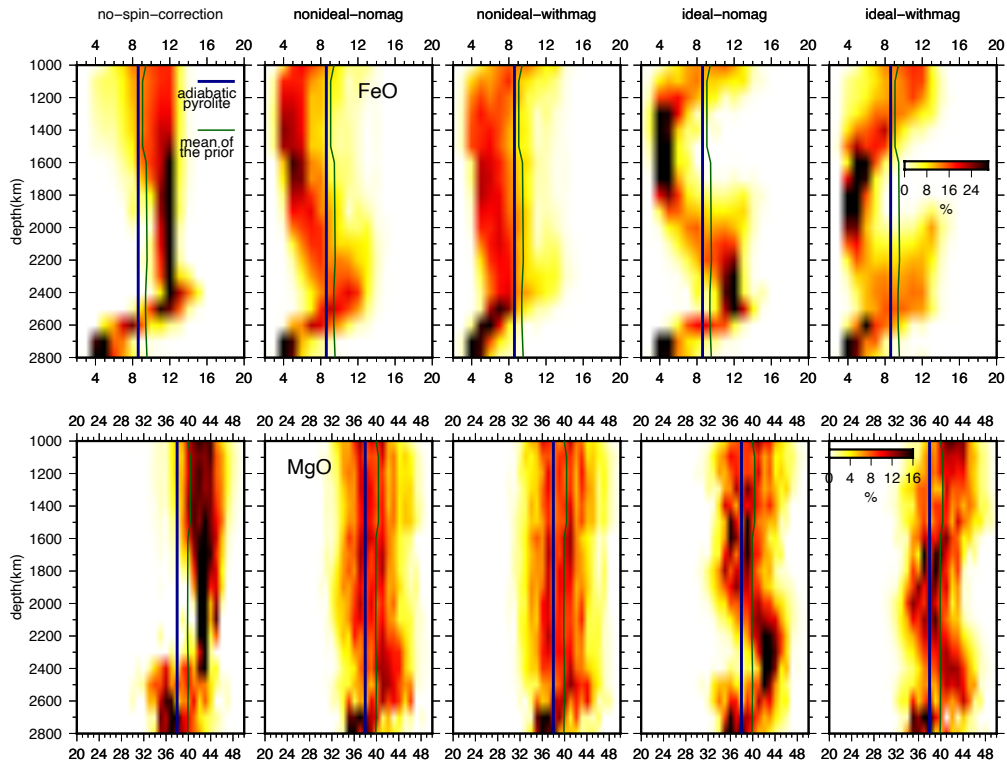

**Fig. S6.**

Density plots showing distributions of wt% FeO (top row) and wt% MgO (bottom row) as a function of depth, for models with variable, but restricted, chemical composition (Prior 3, Fig. S1). Pyrolite (blue line) and mean of the prior (green line) are shown for comparison. On the left, models without a correction to the wavespeeds for spin transition. The other four columns show the results for 4 different spin corrections. Models which include non-ideal solid solution give more reasonable compositional gradients, in particular the model with both non-ideal solid solution and a correction for magnetic entropy. All models show a decrease in iron towards the CMB which is likely driven by the need to increase the bulk wave speed (see main text for discussion).

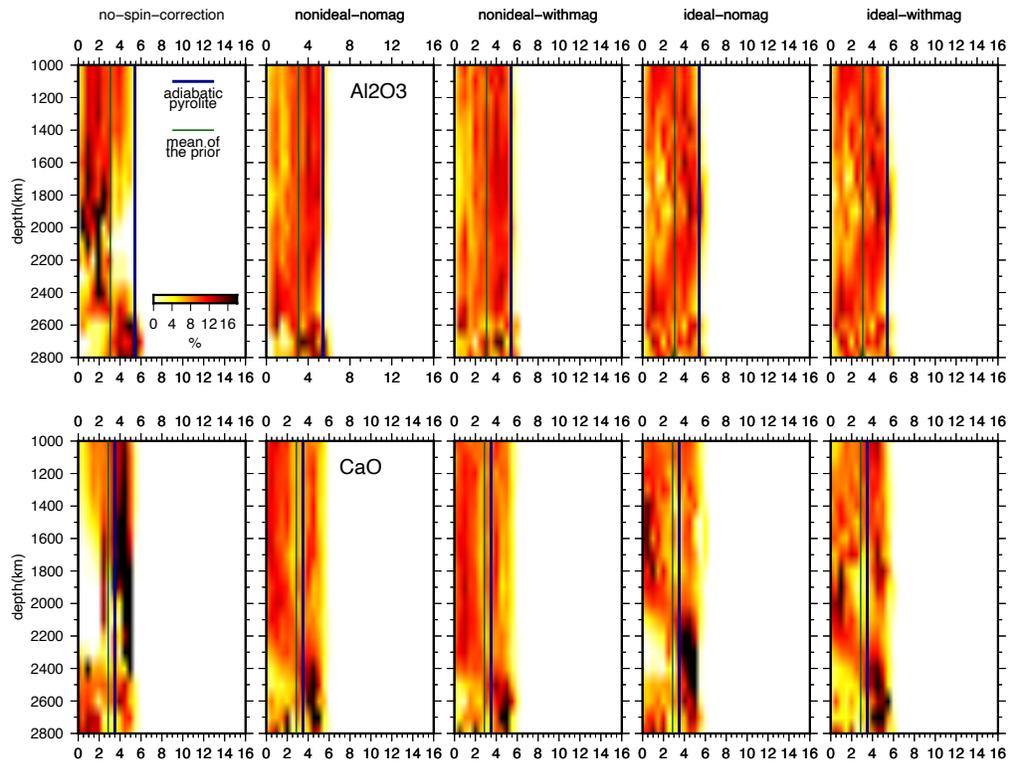

**Fig. S7.**

Density plots showing distributions of wt% Al₂O₃ (top row) and wt% CaO (bottom row) as a function of depth, for models with variable, but restricted, chemical composition (Prior 3, Fig. S1). Pyrolite (blue line) and mean of the prior (green line) are shown for comparison. On the left, models without a correction to the wavespeeds for spin transition. Other four columns show the results for 4 different spin corrections. With non-ideal solid solution the distributions are mostly broad and centred around the mean of the prior.

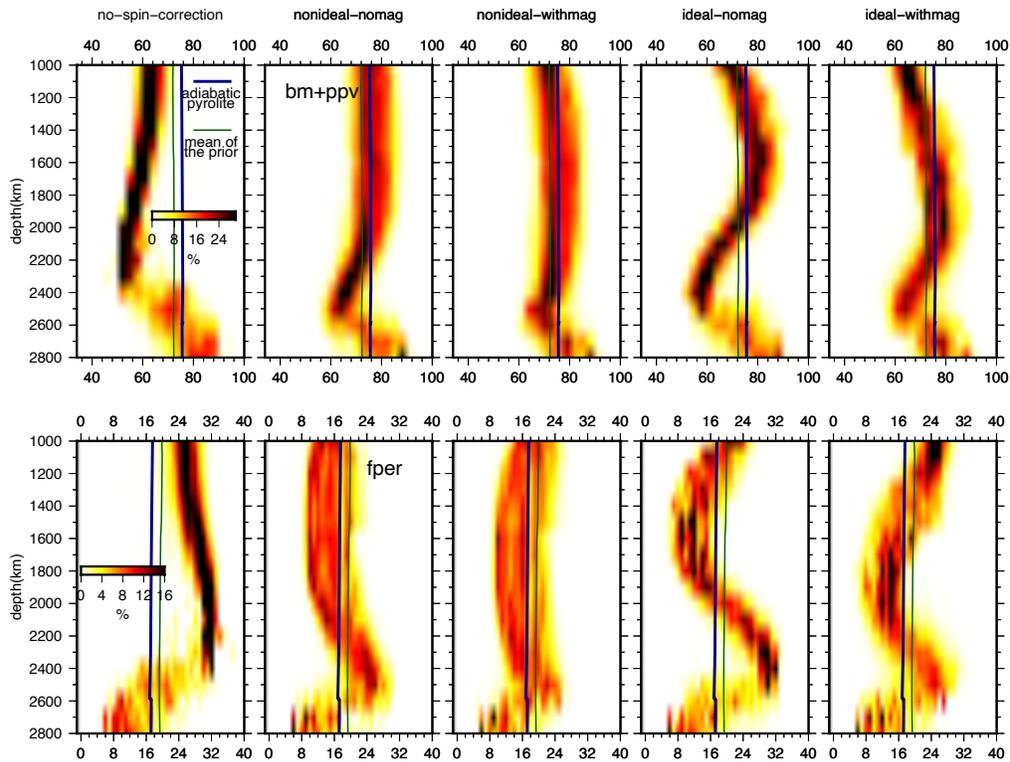

**Fig. S8.**

Density plots showing frequency distributions of minerology as a function of depth, for models with variable, but restricted, chemical composition (Prior 3, Fig. S1). On the top row, (Mg,Fe)SiO₃ (bridgmanite plus post-perovskite), and on the bottom row (Mg,FeO) ferropericlase. Pyrolite (blue line) and mean of the prior (green line) are shown for comparison. On the left, models without a correction to the wavespeeds for spin transition. The other four columns show the results for 4 different spin corrections. Models which include non-ideal solid solution give more reasonable vertical gradients, in particular the model with both non-ideal solid solution and a correction for magnetic entropy. Enrichment in silica towards the CMB is manifested in the mineralogy as an enrichment in (bridgmanite+post-perovskite) and depletion in ferropericlase. Without a correction for spin transition, the bridgmanite content is very low throughout the lower mantle.

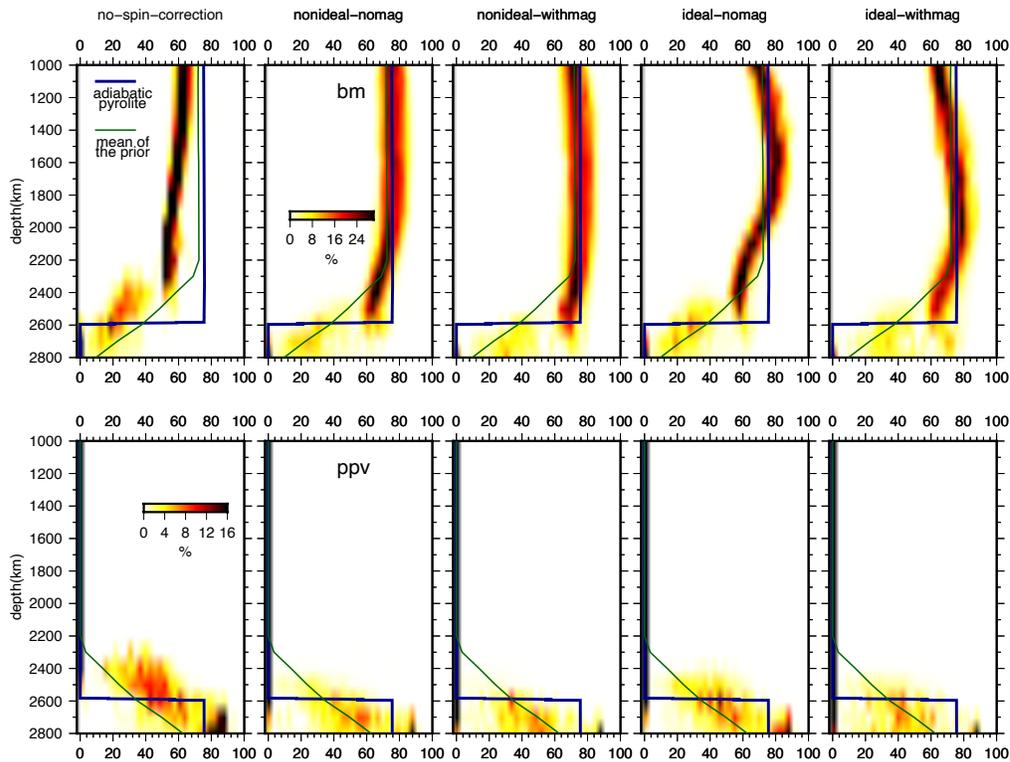

**Fig. S9.**

Density plots showing frequency distributions of minerology as a function of depth, for models with variable, but restricted, chemical composition (Prior 3, Fig. S1). On the top row, (Mg,Fe)SiO$_3$ bridgmanite (bm), and on the bottom row(Mg,Fe)SiO$_3$ post-perovskite (ppv). Pyrolite (blue line) and mean of the prior (green line) are shown for comparison. On the left, models without a correction to the wavespeeds for spin transition. The other four columns show the results for 4 different spin corrections. Without a spin transition, the bridgmanite content is very low throughout the mantle. With a spin transition, the average bridgmanite content is close to pyrolite. Models which include non-ideal solid solution give more reasonable vertical gradients, in particular the model with both non-ideal solid solution and a correction for magnetic entropy.

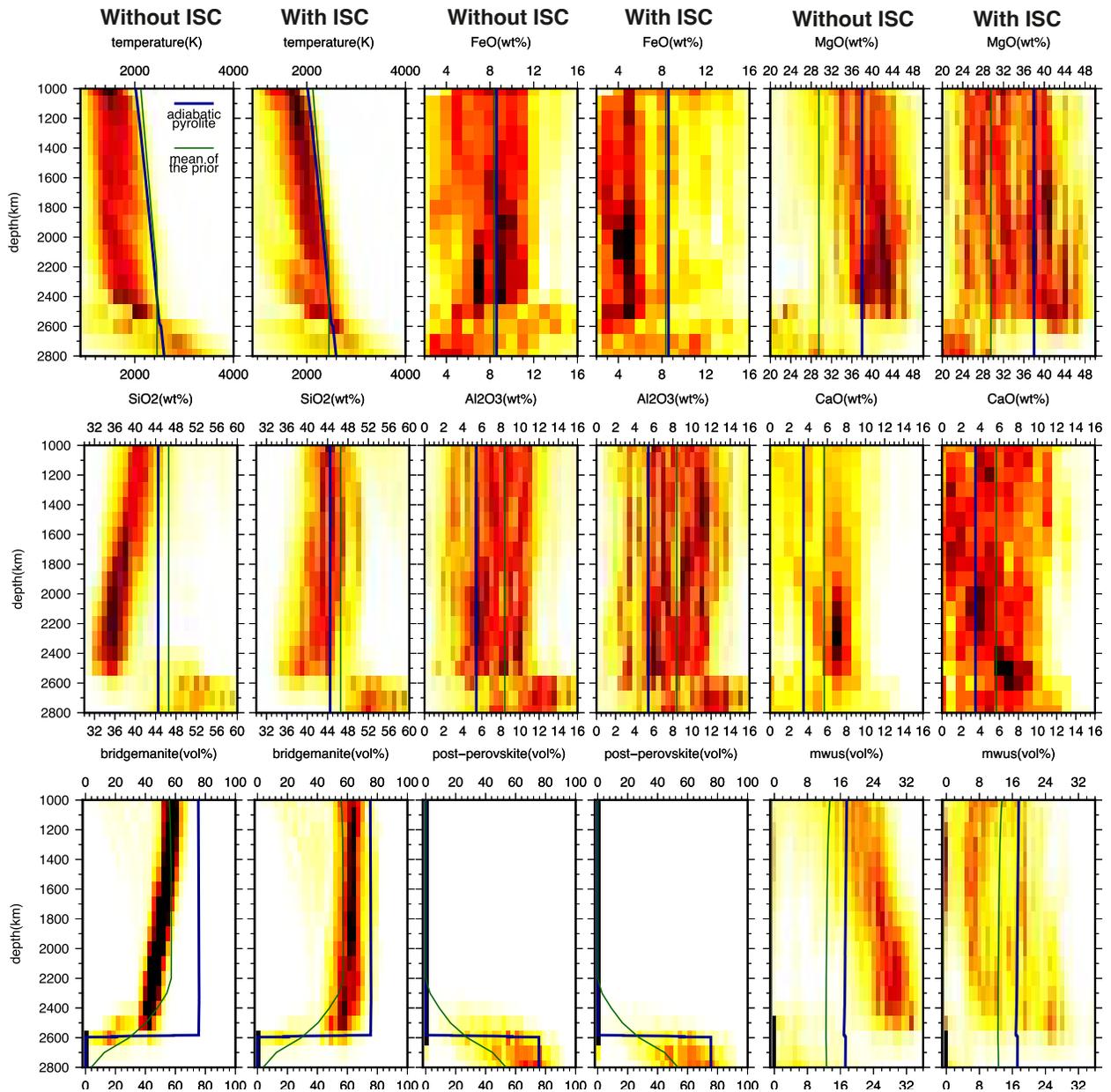

**Fig. S10.**
Density plots showing frequency distributions of temperature, bulk composition and mineralogy (Mg,Fe minerals) for the best-fitting set of models drawn from Prior 2 (Fig S1), as a function of depth. In each pair of plots, the left column is without a correction for spin transition in ferropericlase and the right is with a correction for spin transition. Due to the broad nature of the prior compared to Prior 3 (Figs S5-S9), the posterior distributions are correspondingly broader than those of Prior 3. Models from Prior 3 have bridgmanite contents closer to pyrolitic than Prior 2. The spin correction shown here is with non-ideal solid solution and a correction for magnetic entropy.

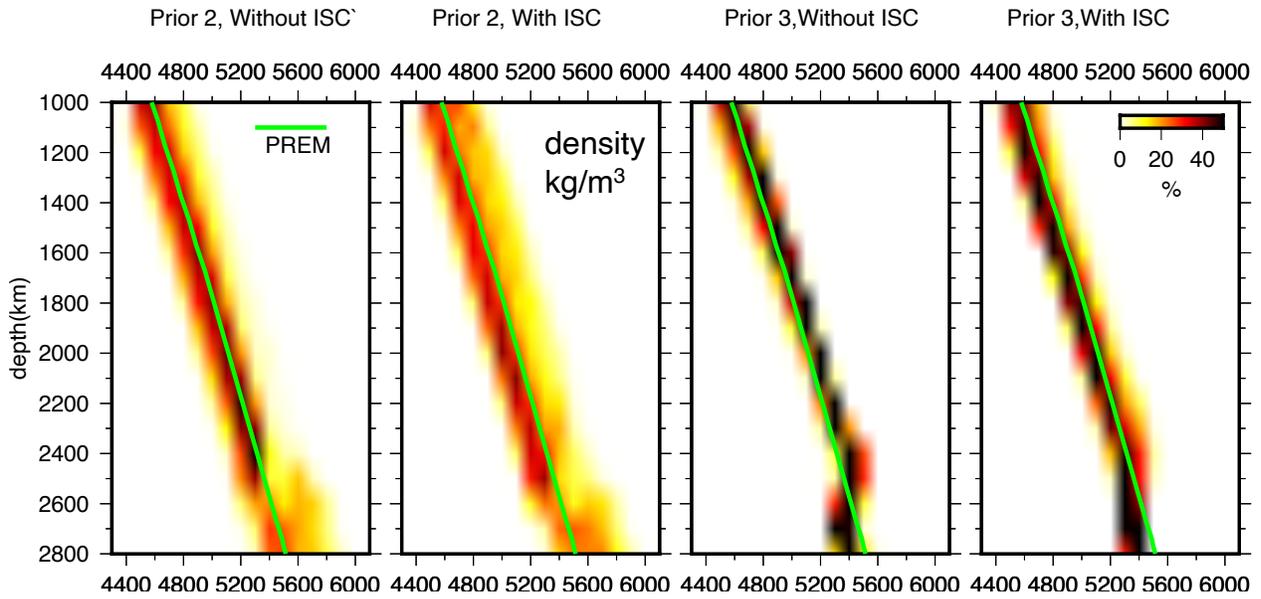

**Fig. S11.**
Frequency distributions of density (horizontal axis, kg/m³) as a function of depth, with and without a spin correction. PREM is shown for reference with a green line. The density distributions correspond to the subset of thermochemical models selected from the stated Prior, by fitting bulk and shear wavespeeds to GLAD-M25. In D", models drawn from Prior 2 (left two plots) follow the trend in PREM more closely than models drawn from Prior 3 (right 2 plots).

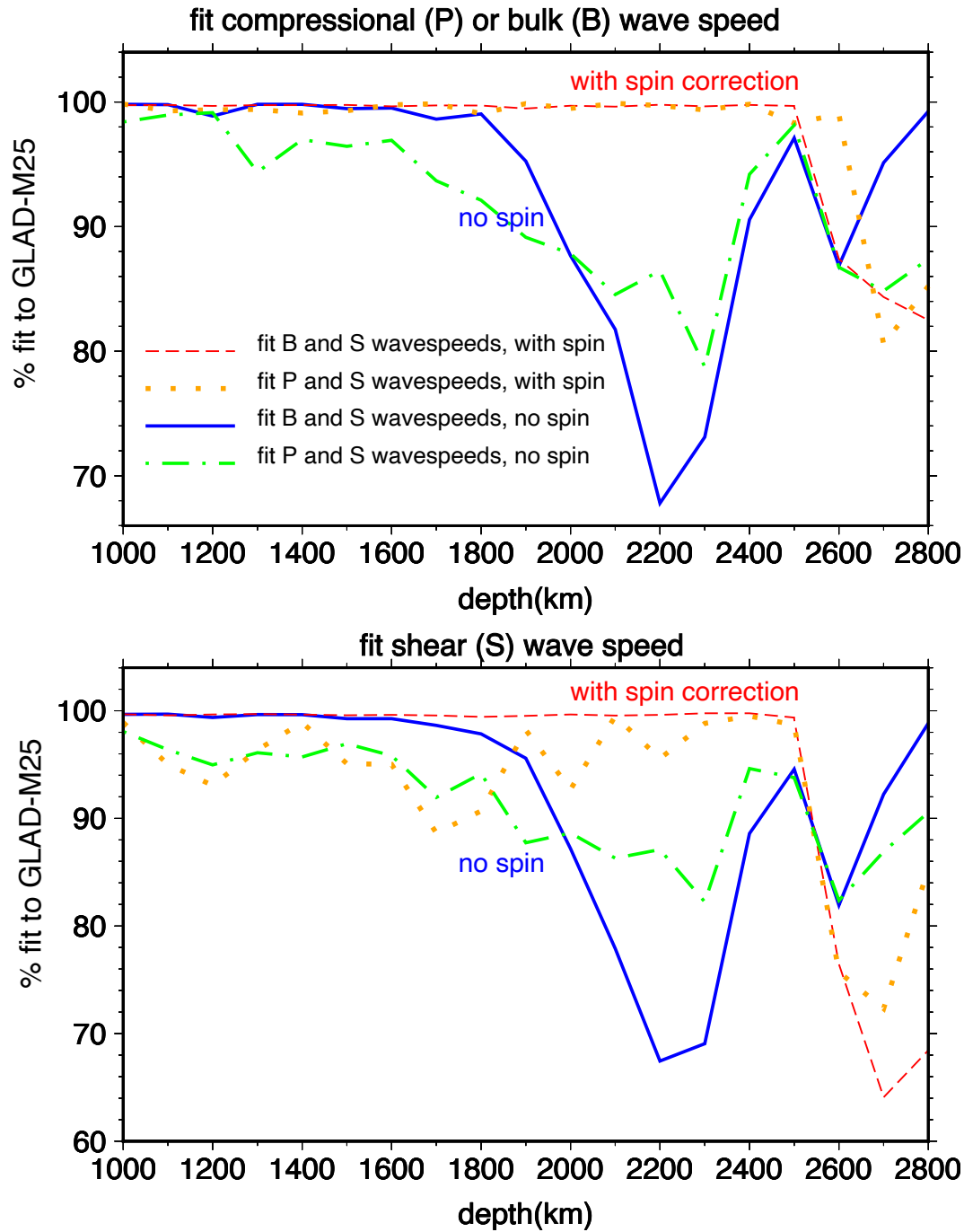

**Fig. S12.**

Comparison of the results when we fit bulk(B) and shear(S) wave-speeds simultaneously (red dashed and blue solid lines) versus fitting compressional (P) and shear wave-speeds simultaneously (orange dotted and green dot-dashed lines). Top panel shows the fit to either $V_P$ or $V_B$ as specified in the legend. Bottom panel shows the fit to $V_S$, having simultaneously fitted either $V_P$ or $V_B$ as specified in the legend. The misfit of NOT including a spin transition in the mid-mantle becomes stronger when we consider bulk wave speed rather than compressional wavespeed (compare not only the blue with the green line, but also the difference between the blue and the red lines, versus the difference between the orange and the green lines).

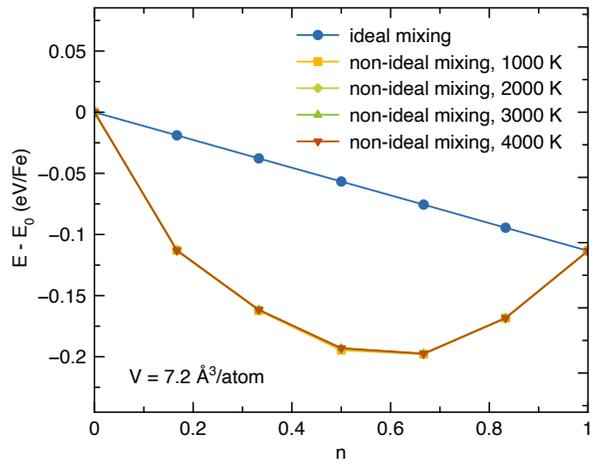

**Fig. S13.**
Static energy $E^{st}(V, T, n)$ per iron vs. $n$ at constant $V$ for $x_{Fe}$=0.1875.

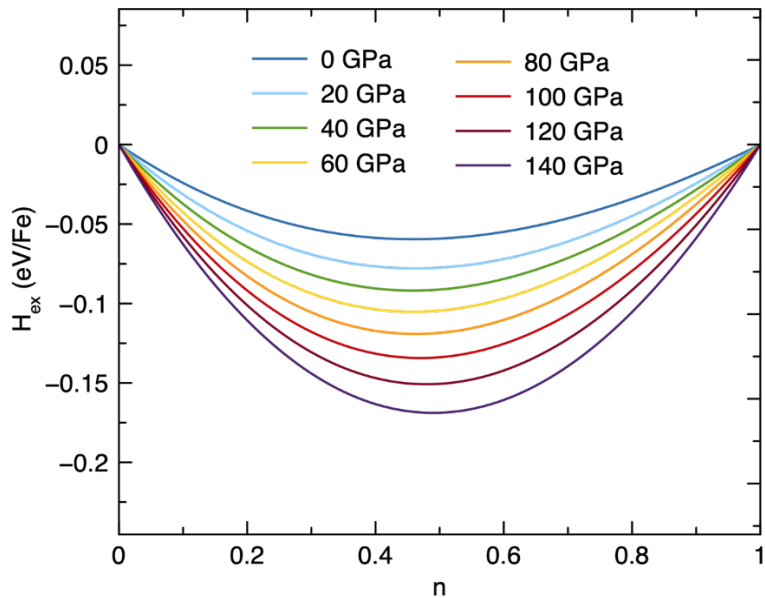

**Fig. S14.**
$H_{ex}(P, n)$ fit to a 3$^{rd}$ order polynomial in $n$ as indicated in Eq. (7).

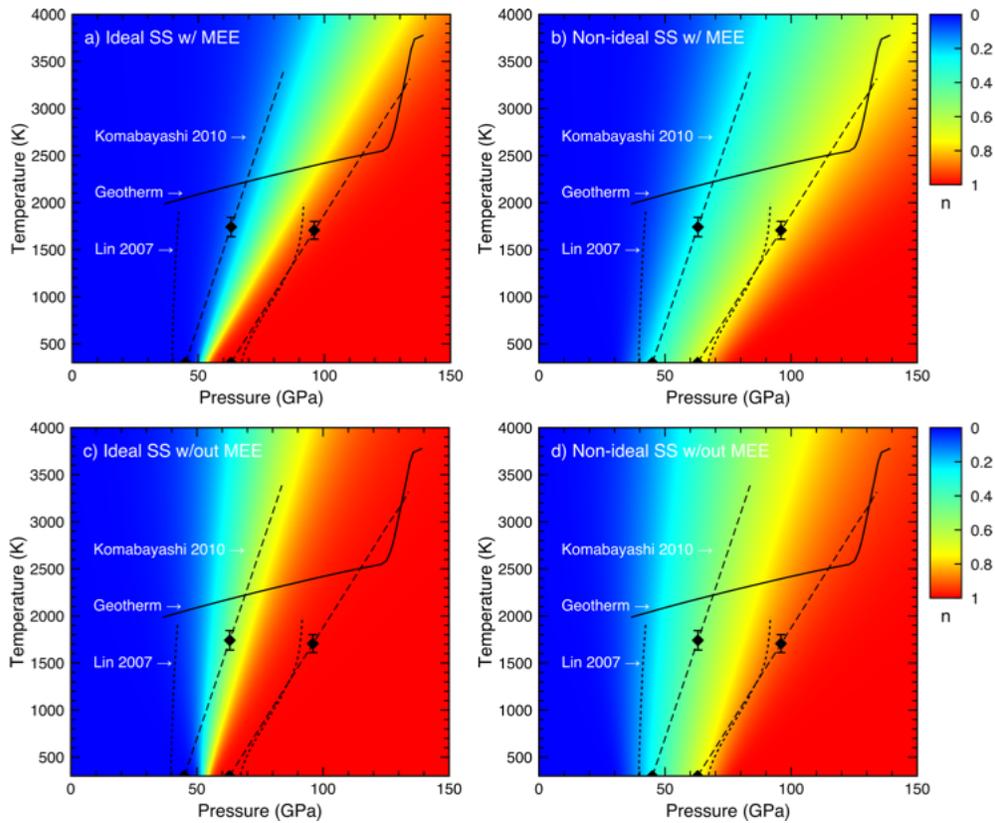

**Fig. S15.**

$n(P, T)$ for x = 0.1875 for four different thermodynamic models of the ISC in *fp*.

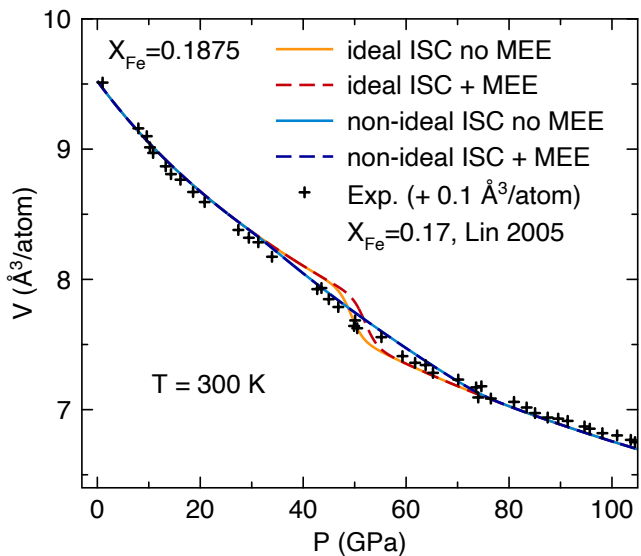

**Fig. S16**

300 K compression curves of the four thermodynamic models of *fp* with x = 0.1875 shown in Fig. S15.

| | | MgO | | SiO2 | | FeO | | Al2O3 | | CaO | | Na2O | |
|---|---|---|---|---|---|---|---|---|---|---|---|---|---|
| | | min | max | min | max | min | max | min | max | min | max | min | max |
| PRIOR 1: PYROLITE | uniform ranges | 37.0 | 39.0 | 44.7 | 46.0 | 7.6 | 10.0 | 3.5 | 4.5 | 3.0 | 3.6 | 0.3 | 0.6 |
| | after normalisation | 36.7 | 39.4 | 44.0 | 46.8 | 7.6 | 10.0 | 3.5 | 4.6 | 3.0 | 3.7 | 0.3 | 0.6 |
| | | | | | | | | | | | | | |
| PRIOR 2: VARIABLE | uniform ranges | 5.0 | 75.0 | 50.0 | 65.0 | 3.0 | 18.0 | 0.0 | 18.0 | 0.0 | 14.0 | 0.0 | 4.0 |
| COMPOSITION (BROAD) | after normalisation | 4.8 | 53.9 | 30.0 | 78.0 | 1.9 | 20.8 | 0.0 | 20.8 | 0.0 | 15.9 | 0.0 | 5.4 |
| | | | | | | | | | | | | | |
| PRIOR 3: VARIABLE | uniform ranges | 33.0 | 52.0 | 42.0 | 51.0 | 4.0 | 15.0 | 0.0 | 6.0 | 0.0 | 6.0 | 0.0 | 0.5 |
| COMPOSITION (BROAD) | after normalisation | 30.5 | 51.5 | 36.3 | 54.3 | 3.6 | 15.8 | 0.0 | 6.3 | 0.0 | 6.4 | 0.0 | 0.6 |

**Table S1.**

Compositional ranges of the three prior distributions plotted in Fig. S1. Oxides are given in wt %. We first draw random values for each oxide from uniform distributions specified by the limits given for "uniform ranges". We then normalize these values so that the sum over the six oxides is 100 %. This has the effect of extending the ranges and making the distributions non-uniform.

| seismic period | 1 | s |
|---|---|---|
| frequency dependence ($\alpha$) | 0.274 | |
| activation energy | 286 | kJ/mol |
| activation volume | $1.2 \times 10^{-6}$ | m$^3$/mol |
| Qref | 312 | |
| Tcore-mantle-boundary | 3500 | kJ/mol |

**Table S2.**

Parameters used in correcting shear wave-speeds for anelasticity, based on (*42*). Seismic period is the period at which GLAD-M25 is calculated.

| Low spin fraction $n$ | Space group of inequivalent configurations | Multiplicity |
|---|---|---|
| $\frac{1}{6}$ | #123 P4/mmm | 1 |
| $\frac{1}{3}$ | #123 P4/mmm | 6 |
| | #131 P4_2/mmc | 6 |
| | #139 I4/mmm | 3 |
| $\frac{1}{2}$ | #123 P4/mmm | 6 |
| | #221 Pm-3m | 2 |
| | #47 Pmmm | 12 |
| $\frac{2}{3}$ | #123 P4/mmm | 6 |
| | #131 P4_2/mmc | 6 |
| | #139 I4/mmm | 3 |
| $\frac{5}{6}$ | #123 P4/mmm | 1 |

**Table S3**. List of inequivalent HS/LS configurations for different values of $n$ and $x_{Fe}$=0.1875.